\documentclass[10pt,conference]{IEEEtran}

\usepackage{graphicx}
\usepackage{url}
\usepackage{amsfonts,amssymb,amsbsy}
\usepackage{amsmath}
\usepackage{tikz}

\usepackage{times}
\usepackage{tikz}
\usetikzlibrary{arrows,automata}

\usetikzlibrary{calc}
\usetikzlibrary{decorations.pathreplacing,decorations.markings,shapes.geometric}
\tikzset{naming/.style={align=center,font=\small}}

\definecolor{royalblue}{cmyk}{1,.5,0,0}
\definecolor{cerulean}{cmyk}{.94,.11,0,0}
\definecolor{violet}{cmyk}{.79,.88,0,0}
\definecolor {Eored}{rgb}{.647,.129,.149}
\definecolor {Eogreen}{rgb}{0,.53,0}

\usetikzlibrary{positioning} \definecolor{processblue}{cmyk}{.96,0,0,0}

\begin{document}

\title{Device-to-Device Data Storage\\with Regenerating Codes}

\author{Joonas P\"a\"akk\"onen$^{\dagger}$, Camilla Hollanti$^{*}$, Olav Tirkkonen$^{\dagger}$
\\
$^\dagger$Department of Communications and Networking, School of Electrical Engineering, \\
$^*$Department of Mathematics and Systems Analysis, School of Science, \\
Aalto University, Espoo, Finland. \\
\{joonas.paakkonen, camilla.hollanti, olav.tirkkonen\}@aalto.fi}

\maketitle

\begin{abstract}

Caching data files directly on mobile user devices combined with device-to-device (D2D) communications has recently been suggested to improve the capacity of wireless networks. We investigate the performance of regenerating codes in terms of the total energy consumption of a cellular network. We show that regenerating codes can offer large performance gains. It turns out that using redundancy against storage node failures is only beneficial if the popularity of the data is between certain thresholds. As our major contribution, we investigate under which circumstances regenerating codes with multiple redundant data fragments outdo uncoded caching.
\end{abstract}

\section{Introduction}\label{introductionsec}

As the amount of mobile data traffic is predicted to keep growing rapidly in the near future \cite{cisco}, more efficient data transmission and distribution methods are needed. Mobile video traffic has quickly become one of the most important factors straining the already burdened cellular networks. As video files are often large, they typically incur significant stress on both cellular networks and backhaul links. Thus, moving traffic away from the traditional cellular and backhaul links could drastically reduce the strain on these links. Further, finding cost-efficient solutions to deliver large, popular data files is important for minimizing the energy consumption of data transmission.

We have observed that the storage space of mobile devices has been increasing. This leads us to the following question: how could we utilize this storage capacity to improve wireless networks? One idea is to use this storage to cache files and distribute them directly between users.

Recently, distributing data directly from devices through device-to-device (D2D) communication has been studied in \cite{han,li,golrezaei}. Principal work on caching as a prefetching method has been conducted in \cite{maddah,bastug}, whereas seminal work on distributed caching, particularly for D2D networks, has been done in \cite{ji}. While coding has been suggested to improve the performance of caching systems \cite{hachem,li2,monteiro,shanmugam}, most of the work in the literature offers no solution to keep the cached files available even when the caching devices move out of coverage.

In this paper, we investigate how redundancy could be used to ensure file availability within a designated area -- even if some nodes \emph{fail}, i.e. leave the area and become unavailable. Namely, we study the performance of regenerating codes \cite{dima} that are codes designed specifically for distributed storage. For further reading, e.g. \cite{tailor} provides an overview of such codes.

We are interested in the performance of the minimum storage regenerating (MSR) and the minimum bandwidth regenerating (MBR) codes, which lie on the far ends of the storage-bandwidth tradeoff curve \cite{dima,rashmi2}. The performance is measured in terms of the expected total transmission cost of the system. Unlike our prior work on similar problems \cite{paakk,paakk2}, the current paper assumes both infinite storage capacities on the users, and that the system must be able to cope with multiple simultaneous failures. That is, even if several users leave the coverage area, the data should still remain available for download from the storage nodes.

We find that the popularity of the file, the number of users, and the transmission costs affect which storage method should be chosen. With the help of numerical results, we characterize the decision rules on choosing the optimal method.

\section{System Model}\label{systemmodelsec}
The current work is based on three key assumptions. Firstly, we assume that mobile user devices have plenty of free storage capacity that can be used to store data. Secondly, we assume that these devices can be used to distribute the stored data to other users via perfect, error-free D2D links. Thirdly, we assume that, on average, transmitting data between mobile devices is less expensive than transmitting data from a base station to a user. This assumption is mostly motivated by the path loss laws of wireless signals, i.e. more transmit power is needed to transmit signals over longer distances. We assume that the average distance between the base station is larger than the average distance between any two nodes.

Based on these assumptions, we show that storing data files with redundancy can lead to significant cost savings. Furthermore, we find explicit thresholds for choosing the most appropriate file storage method given the system parameters.

In our system model, users stay in the system for a random, exponentially distributed amount of time with expected value $T$. We say that the rate at which users pass through the system is $\lambda=\frac{1}{T}$, which can be also thought of as the expected node failure rate.

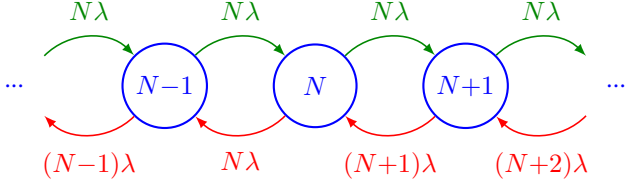
\begin{figure}[tbhp]
\begin{center}
\begin{tikzpicture}
[-latex,auto,auto,node distance=2cm and 2cm,on grid, semithick,state/.style={circle,thick,draw,blue, text=blue,minimum width=1.1cm}]
\node[state] (X) {$N$};
\node[state] (Xm) [left=of X] {$N{-}1$};
\node[state] (Xp) [right=of X] {$N{+}1$};
\node[state,draw=none] (finish) [left of=Xm] {...};
\node[state,draw=none] (end) [right of=Xp] {...};
\path (finish) edge [bend right=-45,color=Eogreen] node[above=0.1cm]{$N\lambda$} (Xm);
\path (Xm) edge [bend right=-45,color=Eogreen] node[above=0.1cm]{$N\lambda$} (X);
\path (X) edge [bend left=45,color=Eogreen] node[above=0.1cm]{$N\lambda$} (Xp);
\path (Xp) edge [bend right=-45,color=Eogreen] node[above=0.1cm]{$N\lambda$} (end);
\path (end) edge [bend right=-45,color=red] node[below=0.1cm]{$(N{+}2)\lambda$} (Xp);
\path (Xp) edge [bend right=-45,color=red] node[below=0.1cm]{$(N{+}1)\lambda$} (X);
\path (X) edge [bend left=45,color=red] node[below=0.1cm]{$N\lambda$} (Xm);
\path (Xm) edge [bend right=-45,color=red] node[below=0.1cm]{$(N{-}1)\lambda$} (finish);
\end{tikzpicture}
\end{center}
\caption{M/M/$\infty$ Markov chain state diagram for the instantaneous number of nodes (blue). The incoming rate (green) of the nodes is constant, whereas the outgoing rate (red) is proportional to the number of nodes in the system. The expected number of nodes is $N$ and $\lambda=1/T$.}
\label{mchain}
\end{figure}

We denote the expected number of nodes in the system by $N$. We assume that the instantaneous number of nodes can be described by the M/M/$\infty$ Markov model, shown in Fig. \ref{mchain}, where the state corresponds to the instantaneous number of nodes. It is well-known that the probability that this chain is in state $i$ is \cite{harrison}
\begin{align}
\pi(i)=\frac{N^i}{i!}e^{-N}.
\label{poisson}
\end{align}

% THIS IS THE MARKOV CHAIN:

Without loss of generality, let there be one file of size $B=1$. Let us assume that each user that is connected to the system requests the file at random, exponentially distributed time intervals with expected value $\tau=\frac{1}{\omega}$, where $\omega$ is called the \emph{file request rate}.

We assume that files are always available, either from the base station or from a set of storage nodes. Let $R>1$ denote the expected cost ratio between transmitting a bit from the base station and transmitting a bit from another user through a D2D link. That is, the cost of retrieving the file from the base station is $R$, while the cost of retrieving the file from another user is only $1$. Note that $R$ could be either based on measurements, or it could be artificially set by the system designer to adjust the amount of traffic offloaded from the base station to the D2D connections. The higher the value of $R$, the more traffic is moved away from the base station.

Additionally, let $p=\omega T=\frac{\omega}{\lambda}$ be the expected number of requests that one user generates during the time it spends in the system. As it is reasonable to assume that users do not generally request a certain file more than once during their visit to the system, we mainly focus on the case $p<1$.

\begin{figure}[tbhp]
\centering \includegraphics[scale=.5]{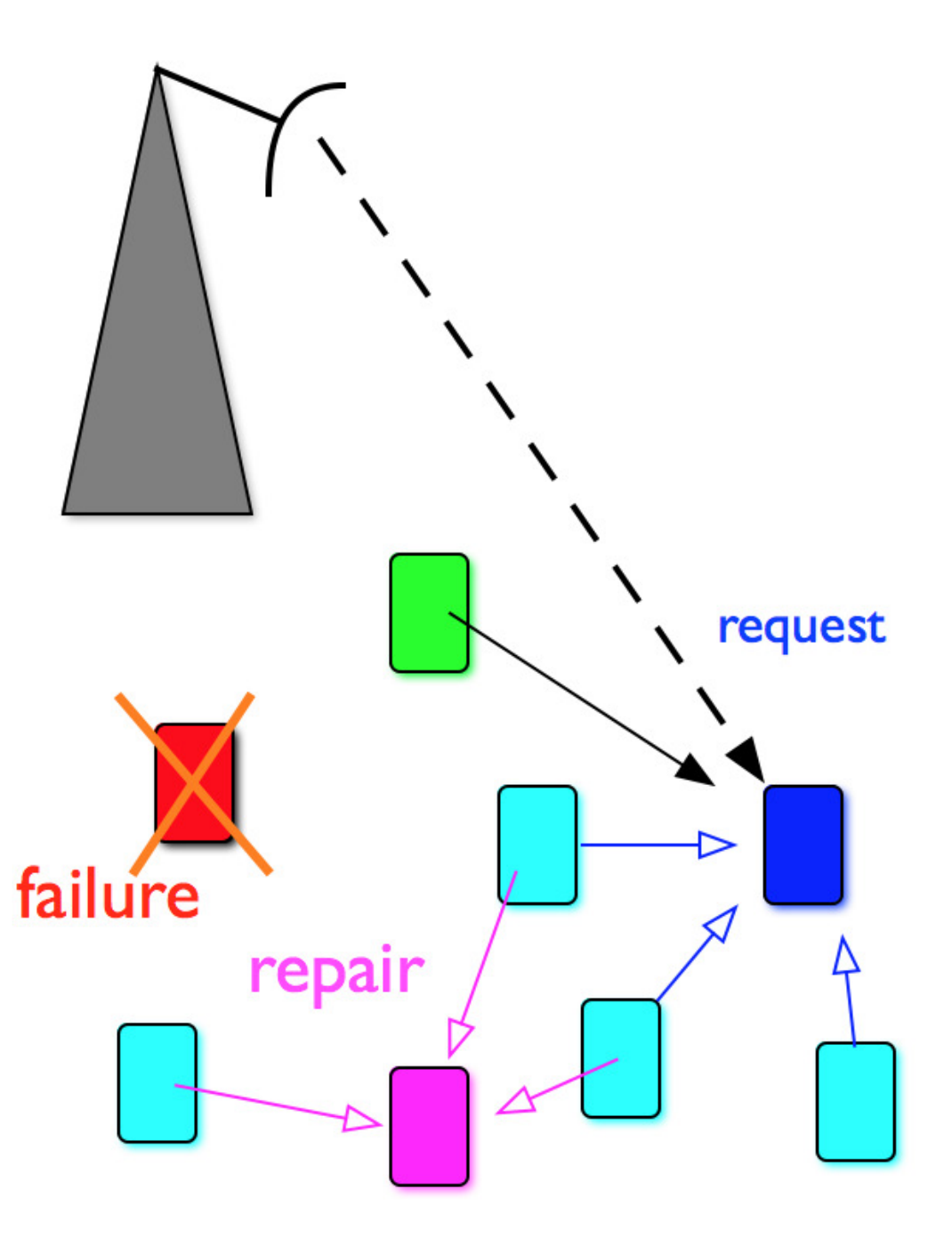}
\caption{Node requesting (blue) a file can be served by the base station (grey), a single storage node (green, simple caching or replication), or by a set of storage nodes each transmitting an encoded data block. When a node fails (red), the lost block can be repaired to a new node (magenta). Here $d=k=3$.}
\label{newgreatbasestation}
\end{figure}

Fig. \ref{newgreatbasestation} illustrates the system setup and the select data storage and distribution methods along with the repair process, which we discuss in more detail in the following section.

\section{Analysis}\label{analysissec}
In this section, we introduce the storage methods simple caching, regenerating codes and replication. We derive closed-form expressions for the expected total cost per time unit for each of them. We note that traditional erasure coding and retrieving data directly from the base station cannot outperform MSR and simple caching, respectively. For this reason, we do not consider these two methods. This is further justified in detail later in this section.

\subsection{Simple caching}
We call the method of storing one full copy of the data file on a single node with no redundancy \emph{simple caching}. As long as the node that is caching the file stays in the
system, all file requests lead to retrievals from this node. There
are, on average, $(N-1)$  nodes that generate requests as the node storing the file does not request the file. Therefore, the expected number of requests during the lifetime of the
caching node is $(N-1)p$.

If the caching node fails, the next node that requests the file has to
download it from the base station. The expected time in which this
happens is $\frac{1}{N\omega}$ as the expected total request rate is
$N\omega$. Therefore, the expected time in which a number of $(N-1)\omega
T+1$ requests are generated is $T+\frac{1}{N\omega}$. The expected
cost of these requests is $(N-1)\omega T+R$ and, thereby, the expected
cost of simple caching becomes
\begin{align*}
C_{\text{sc}} = \frac{(N-1)\omega T + R}{T + \frac{1}{N\omega}} = \frac{(N-1)\omega + R\lambda}{1 + \frac{\lambda}{N\omega}}.
\end{align*}
It should be noted that if we only serve file requests from the base station, the expected cost becomes $RN\omega$. It is easy to see that this method cannot beat simple caching, i.e. $C_\text{sc}<RN\omega$, for all $R>1$. This is due to the fact that part of the requests of simple caching are served by a cheaper D2D connection. Thus, we do not consider the method of serving users only via the base station.

\subsection{Redundant caching with regenerating codes}
Here we use regenerating codes \cite{dima} to ensure file availability. Regenerating codes with parameters ($n,k,d$) are maximum distance separable (MDS) codes that allow any $k$ nodes to be contacted to recover the file. Furthermore, regenerating codes possess the so called \emph{reconstruction property}, which says that contacting any $d$ nodes allows resurrecting a lost node. Throughout this work, we call $k$ the reconstruction degree and $d$ the repair degree.

There are two extreme cases of regenerating codes: the minimum storage regenerating (MSR) code and the minimum bandwidth regenerating (MBR) code. For example \cite{rashmi2} provides code constructions for both the MBR and the MSR point. The MSR code minimizes the number of data stored on the storage nodes, while the MBR code minimizes the amount of traffic required when repairing a lost data block. Here the amount of information stored on each node is denoted $\alpha$, and the amount of information communicated at each repair is denoted $\gamma$. In \cite{dima}, the values of $\alpha$ and $\gamma$ for MBR and MSR were derived to yield

\begin{align}\label{MBRformula}
(\alpha_{\text{MBR}},\gamma_{\text{MBR}}) = \left(\frac{2Bd}{2kd-k^2+k},\frac{2Bd}{2kd-k^2+k}\right)
\end{align}

\begin{align}\label{MSRformula}
(\alpha_{\text{MSR}},\gamma_{\text{MSR}}) = \left(\frac{B}{k},\frac{Bd}{k(d-k+1)}\right),
\end{align}
where $B$ is the file size, which we set to $B=1$ in this work without loss of generality.

It should be noted that the MSR code with $d=k$ is equivalent to traditional MDS erasure coding. Furthermore, when $d>k$, MSR outperforms traditional MDS coding because of its lower repair bandwidth. Thus, we do not consider traditional erasure coding as a separate coding method in this work.

\begin{figure}[tbhp]
\centering \includegraphics[scale=.31]{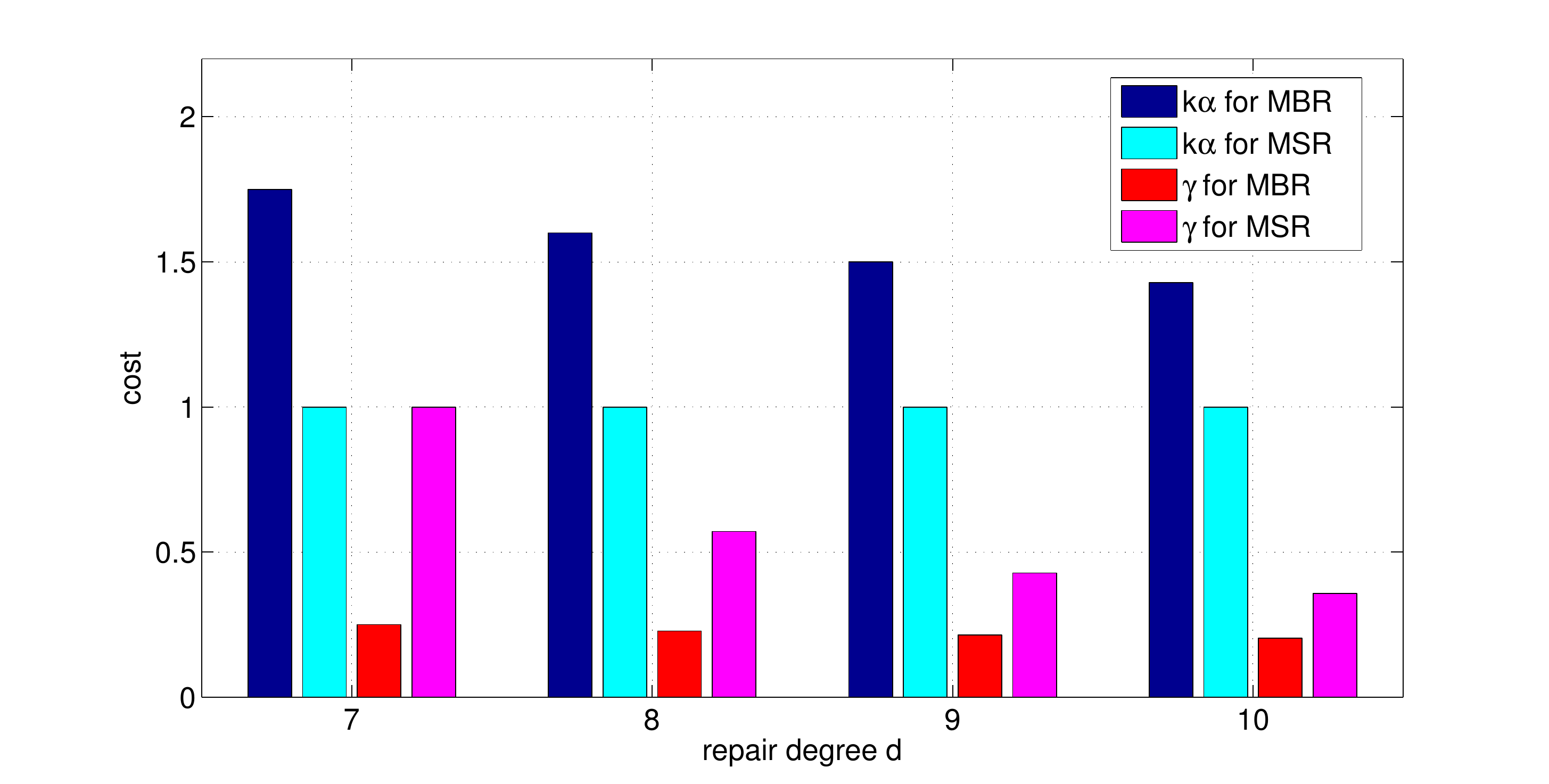}
\caption{Example of file reconstruction and repair costs for $k=7$ and $d\in \{7,8,9,10\}$. The reconstruction cost $k\alpha$ of MBR (blue) is higher than that of MSR (cyan), while the repair cost $\gamma$ is lower for MBR (red) than for MSR (magenta). Also observe that the highest value of $d$ yields the lowest costs.}
\label{tradeoff}
\end{figure}

Even though the MBR code minimizes the amount of traffic required when a node becomes unavailable and its contents must be regenerated to another node, the storage space needed for MBR is higher than that of MSR. In view of the current work, more importantly, the reconstruction bandwidth is higher for MBR than for MSR. That is, MBR requires more information than the size of the file to be transmitted every time a user requests the file\footnote{It is important to note that we assume that the reconstructing node always downloads all the $\alpha$ symbols from the $k$ storage nodes to which it connects.}. Therefore, whether to apply MBR or MSR, or either, largely depends on the time the users spend in the system, and the popularity of the file. Fig. \ref{tradeoff} shows the tradeoff between reconstruction bandwidth ($k\alpha$) and repair bandwidth ($\gamma$) for certain code parameters. Note that there exist also regenerating codes that offer a tradeoff between MBR and MSR. However, we do not consider these codes in this work for the sake of simplicity.

Now we derive the exact expression for the cost function for regenerating codes as a function of parameters $R,N,\omega,\lambda,n,k$ and $d$. These expressions are general in terms of the repair bandwidth $\gamma$ and the size of the stored block $\alpha$. Thereby, the expressions can be used for both the MSR and the MBR code -- only the values of $\gamma$ and $\alpha$ must be changed.

We divide the expected total cost expression of regenerating codes into six costs: allocation cost $C_1$, cost of creating redundancy $C_2$, repair cost $C_3$, cost of remote retrievals $C_4$, cost of data reconstruction by storage nodes $C_5$, and cost of data reconstruction with many nodes $C_6$. In the following, we further explain these cost terms and present the expected cost of each term.

% Allocation cost C1
\emph{Allocation cost:} Each time there are exactly $k-1$ nodes and a new node enters the system, which happens with probability $\pi(k-1) \frac{N}{k-1+N}$, the base station allocates a block of size $\alpha$ to all $k$ nodes. Note that, to get the expected cost over time, this cost must be normalized by the expected time that a user spends in the system, which is simply $\sum_{i=0}^{\infty} \left( \frac{T}{i+N}\right) \pi(i) = \frac{1}{2N\lambda}$. Thus, the expected cost of reallocation after data loss becomes
\begin{align*}
C_1 = 2N\lambda \pi(k-1) \left(\frac{N}{k-1+N}\right)Rk\alpha.
\end{align*}

% Cost of creating redundancy C2
\emph{Cost of creating redundancy:} This process creates the desired redundant data blocks. If the number of nodes is in $[k,d-1]$ when a new node appears, we transmit $k\alpha$ bits to the new node, while if the number of nodes is in $[d,n-1]$, we only need to communicate $\gamma$ bits. This cost becomes
\begin{align*}
C_2 = 2N\lambda \sum_{i=k}^{d-1} \pi(i) \left(\frac{N}{i+N}\right)k\alpha + 2N\lambda \sum_{i=d}^{n-1} \pi(i) \left(\frac{N}{i+N}\right)\gamma.
\end{align*}

% Repair cost C3
\emph{Repair cost:} Every time a storage node leaves the system, the system attempts to repair the lost block of data in order to keep the number of stored blocks constant. The probability that there are $i$ nodes, and that the next event is a node departure, and that the departed node was storing a block is $\pi(i)\frac{i}{i+N}\frac{n}{i}$. Repairing is only possible if there is at least one empty node after the departure of a storage node. Thus, we sum over $i\in [n+2,\infty)$. The cost of each repair is $\gamma$, so the repair cost becomes
\begin{align*}
C_3 &= 2N\lambda \sum_{i=n+2}^{\infty} \pi(i) \left(\frac{i}{i+N}\right)\left(\frac{n}{i}\right)\gamma \\
&= 2N\lambda \sum_{i=n+2}^{\infty} \pi(i)\frac{n\gamma}{i+N}.
\end{align*}

% Cost of remote retrievals C4
\emph{Cost of remote retrievals:} If there are fewer than $k$ nodes, the base station must be contacted to download the file. This cost becomes
\begin{align*}
C_4 = \sum_{i=1}^{k-1} \pi(i) i w R.
\end{align*}

% Cost of data reconstruction by storage nodes C5
\emph{Cost of reconstruction by storage nodes:} If the number of nodes is in $[k,n]$, every time a node requests a file, it only needs to connect to $k-1$ other nodes since it already has one block stored on itself. Thus, this cost becomes
\begin{align*}
C_5 = \sum_{i=k}^n \pi(i) i \omega (k-1) \alpha.
\end{align*}

% Cost of data reconstruction with many nodes
\emph{Cost of data reconstruction with many nodes:} If there are more than $n$ nodes, the $n$ nodes that are already storing a block only need to connect to $k-1$ nodes for reconstruction, while the nodes that are not storing anything must connect to $k$ nodes. The cost of these requests becomes
\begin{align*}
C_6 &= \sum_{i=n+1}^{\infty} \pi(i) n \omega (k-1) \alpha + \sum_{i=n+1}^{\infty} \pi(i) (i-n) \omega k \alpha \\
&= \sum_{i=n+1}^{\infty} \pi(i) (ki-n) \alpha \omega.
\end{align*}
Note that, although not shown in the above equations, $\alpha$ and $\gamma$ are functions of $k$ and $d$, just like in \eqref{MBRformula} and \eqref{MSRformula}.

The performance metric in which we are interested, i.e. the expected total cost, becomes the sum of all the above six costs. However, if the average number of nodes is much higher than the average number of nodes storing a data block, i.e. if $N\gg n$, only the repair cost $C_3$ and the reconstruction cost with many nodes $C_6$ count since all the other events become extremely rare. Nevertheless, in the numerical results of this work, we take all the six events into consideration.

\subsection{Replication}
When replication is used, $n$ nodes store an exact replica of the data file. If we set $k=\alpha=\gamma=1$, we can use the sum of all the six expressions of regenerating codes in the previous section to find the cost of the replication method. While replication is simple and has a minimum reconstruction bandwidth, its drawback is its high repair bandwidth. Additionally, replication consumes plenty of storage space. This, however, is not important here as we assume that all nodes have very large storage capacities.

We point out that the expressions for the cost of simple caching, caching with regenerating codes, and replication could be used to analytically find the best method for given system parameters. Due to the laborious nature of this task and the lack of space, however, we only find the optimal methods with the help of numerical computations. Additionally, it is important to note that finding the optimal method analytically only yields inequations of $p=\frac{\omega}{\lambda}$, i.e. only the ratio of $\omega$ and $\lambda$ matters, not the actual values.

\section{Numerical Results}\label{numericalresultssec}
It may be desirable that the number of storage nodes that participate in the repair and reconstruction processes in a distributed storage system is high because high reconstruction and repair degrees imply low transmission costs. However, for our system setup, we intentionally keep the number of participating nodes relatively low. We limit the values of parameters $k$ and $d$ to a certain maximum. This is because, in practice, it can be very difficult to establish a large number of simultaneous D2D links whenever a user wants to reconstruct the file, or when a failed node must be repaired.

Setting up several parallel data streams could speed up the download process, which would motivate keeping $k$ and $d$ relatively large. The faster the D2D link to a certain node is, the more data could be retrieved from that node. However, both parallel and asymmetric downloads are outside of the scope of this paper, but they could be investigated in future work.

For the numerical analysis of this section, we set the maximum repair degree to $d=10$. More precisely, we will always use $d=10$ as it is obvious from \eqref{MBRformula} and \eqref{MSRformula} that maximizing $d$ minimizes both $\alpha$ and $\gamma$. Further, fixing $d=10$ implies that $k\in[2,10]$, as $2\leq k\leq d$.

Even though here it is sensible to limit the values of $k$ and $d$, it is beneficial to keep the number of storage nodes $n$ relatively high. Here we set $n=30$, which we consider to be high enough to avoid losing the file too easily due to potential multiple simultaneous failures, but still low enough so that we can assume that the average number of nodes is much less than the desired number of storage nodes, i.e. $n\ll N$. In practice, the value of $n$ would affect the data transmission cost ratio $R$. If the downloading node can choose the $k$ or $d$ closest nodes to contact, it would always be beneficial to have $n$ as high as possible. However, in this work, we ignore this effect due to its complex nature and note that this could be another direction of future work.

\subsection{Finding the optimal method}
\begin{figure}[tbhp]
\centering \includegraphics[scale=.4]{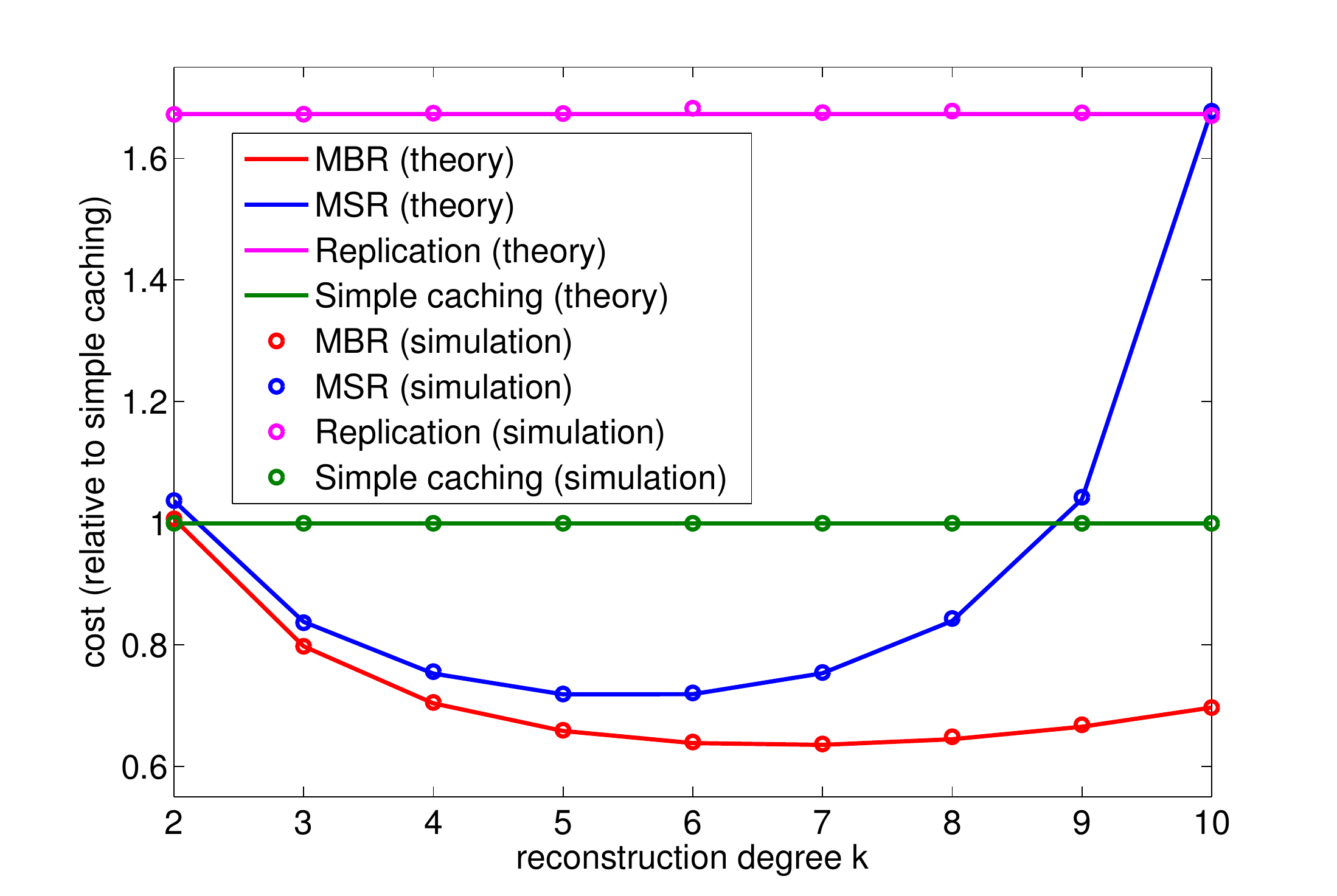}
\caption{Costs of regenerating codes relative to simple caching as functions of $k$ with $R=20$, $N=1000$, and $p=0.005\approx 10^{-2.3}$ (cf. Fig. \ref{costaspfig}). Here MBR with $k=7$ yields the best performance. Simple caching and replication are independent of $k$, but the simulations are repeated for each point on the lines.}
\label{costsask2fig}
\end{figure}

Finding the method that yields the minimum cost is rather complicated because of the large number of both system parameters and code parameters. We need to compare the cost of simple caching and replication to the minimum costs of MBR and MSR. Fig. \ref{costsask2fig} suggests that finding the optimal $k$ is not trivial. The same figure shows that the gains are notable, especially when using the MBR code with $k=7$ in this example.

\begin{figure}[tbhp]
\centering \includegraphics[scale=.4]{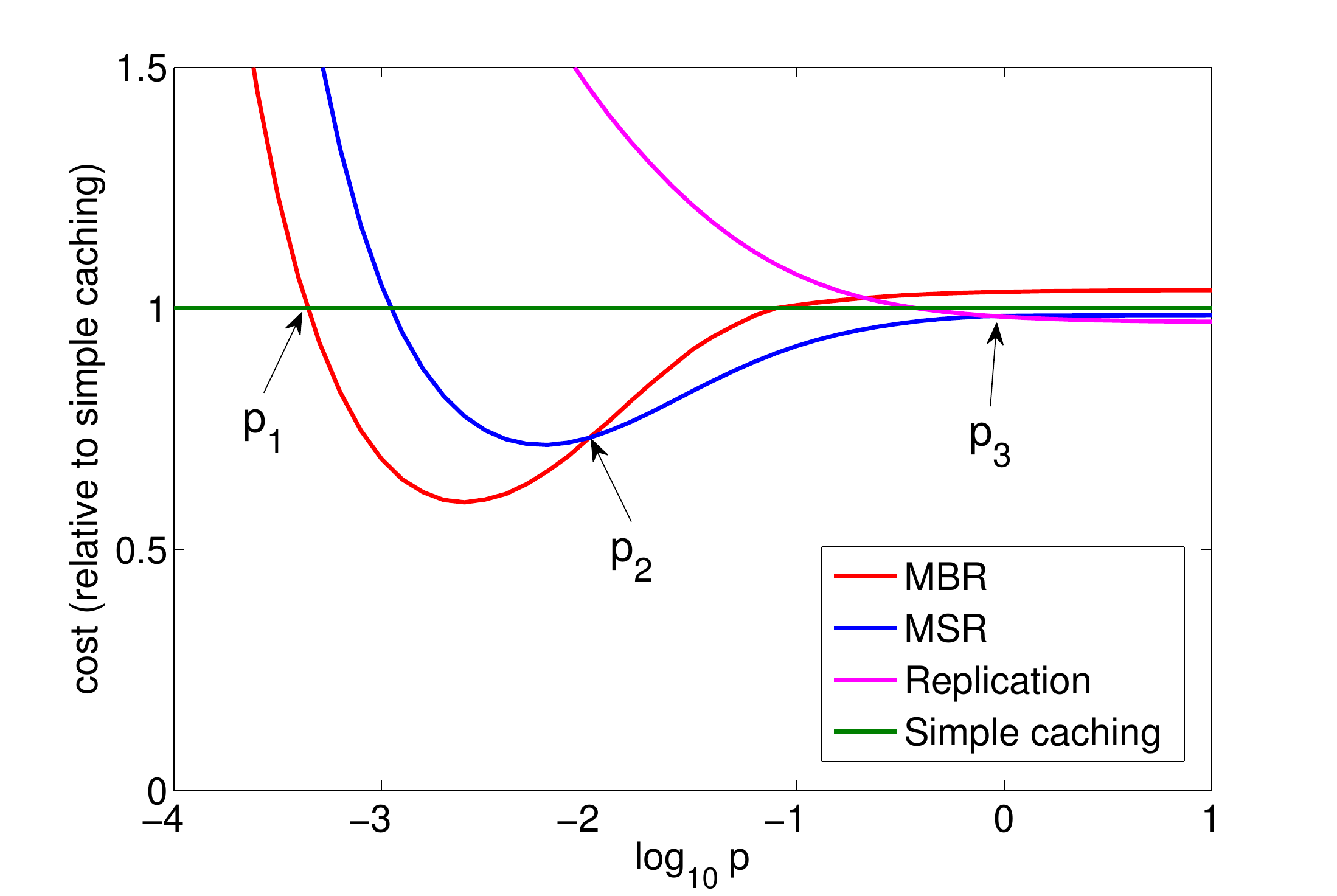}
\caption{Costs as a function of the expected number of file requests during the lifetime of a node ($p$) for $R=20$ and $N=1000$. Here both MBR and MSR can yield significant savings, while replication only offers modest improvements for very high request rates. The arrows point to the crossing points of the corresponding curves, i.e. the switching thresholds for $p$.}
\label{costaspfig}
\end{figure}

\begin{figure}[tbhp]
\centering \includegraphics[scale=.4]{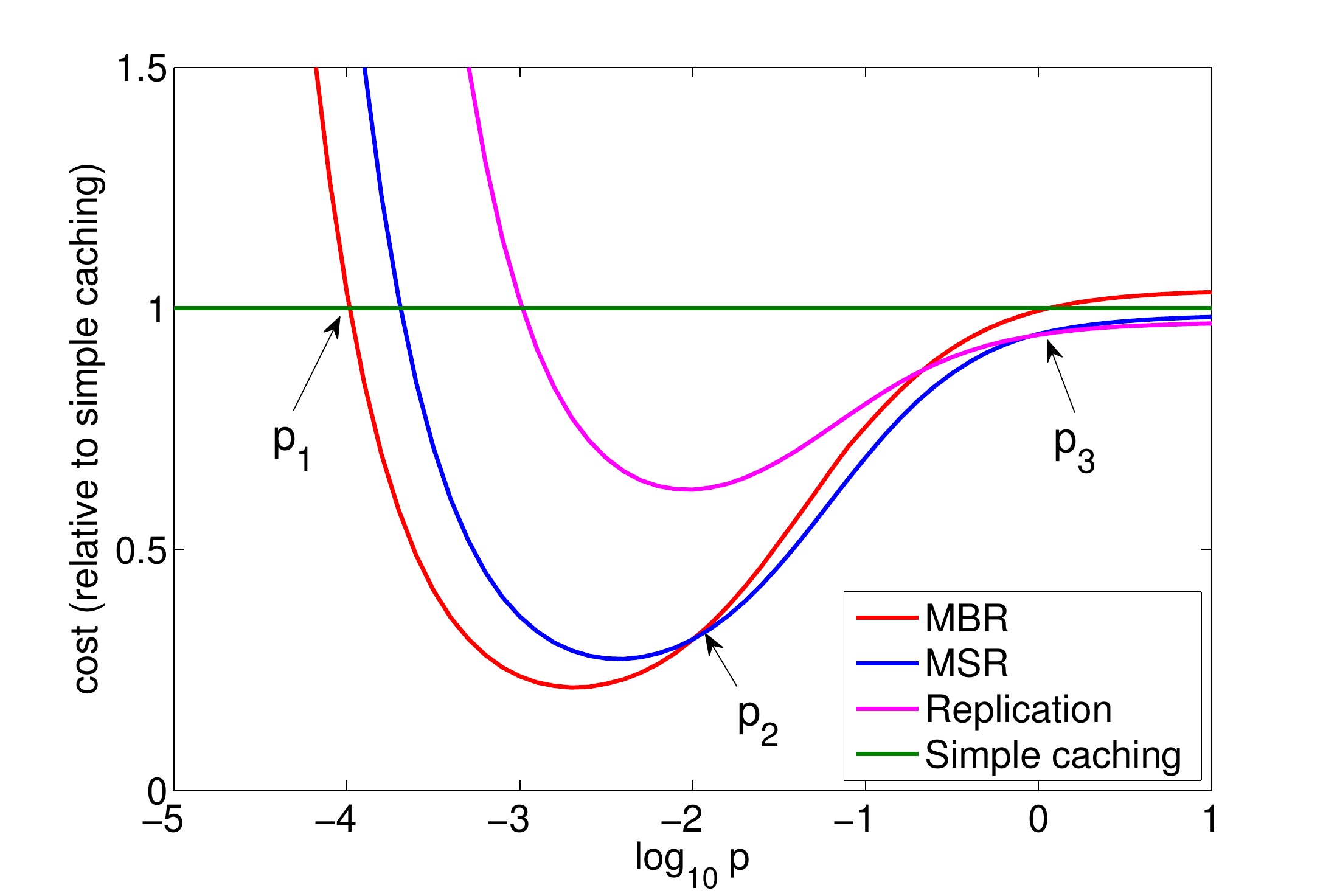}
\caption{Costs as a function of the expected number of file requests during the lifetime of a node ($p$) for $R=60$ and $N=1000$. The cost savings are higher compared to Fig. \ref{costaspfig} as $R$ is higher. The arrows point to the crossing points of the corresponding curves, i.e. the switching thresholds for $p$.}
\label{costasp2fig}
\end{figure}

Figures \ref{costaspfig} and  \ref{costasp2fig} show the optimal performances of each method as functions of $p=\frac{\omega}{\lambda}$. For extremely low values of $p$, the number of failures is large compared to the number of file requests. Therefore, the repair cost vastly dominates the total cost, and it is not worth repairing the file if the request rate is too low. Consequently, simple caching is the desired method here. One might also argue that, in the case of a very low request rate, caching would imply such small cost savings that it should not be used at all.

For higher values of $p$, the number of file requests justifies the use of distributed storage on the nodes but repairs still dominate the total cost. Therefore, MBR performs best in this case. However, further increasing $p$ means that it becomes more and more important to keep the file reconstruction cost low, thus, MSR should be chosen.

For a very high $p$, reconstructions dominate the total cost. Even though the reconstruction cost of MSR is equal to that of replication, replication performs better as it reduces the total request rate. When MSR is used, even the storage nodes that are storing data must download the remaining $k-1$ blocks in order to recover the data file. On the contrary, when replication is used, all of the $n$ storage nodes are already storing the file, so they do not need to download anything.

\subsection{Switching thresholds}
In the remainder of this section we present switching thresholds for certain parameter values. For example, the switching threshold $p_1$ for choosing MBR over simple caching means that if $p>p_1$, then MBR should be chosen over simple caching because it yields a lower expected total cost.

\begin{figure}[tbhp]
\centering \includegraphics[scale=.41]{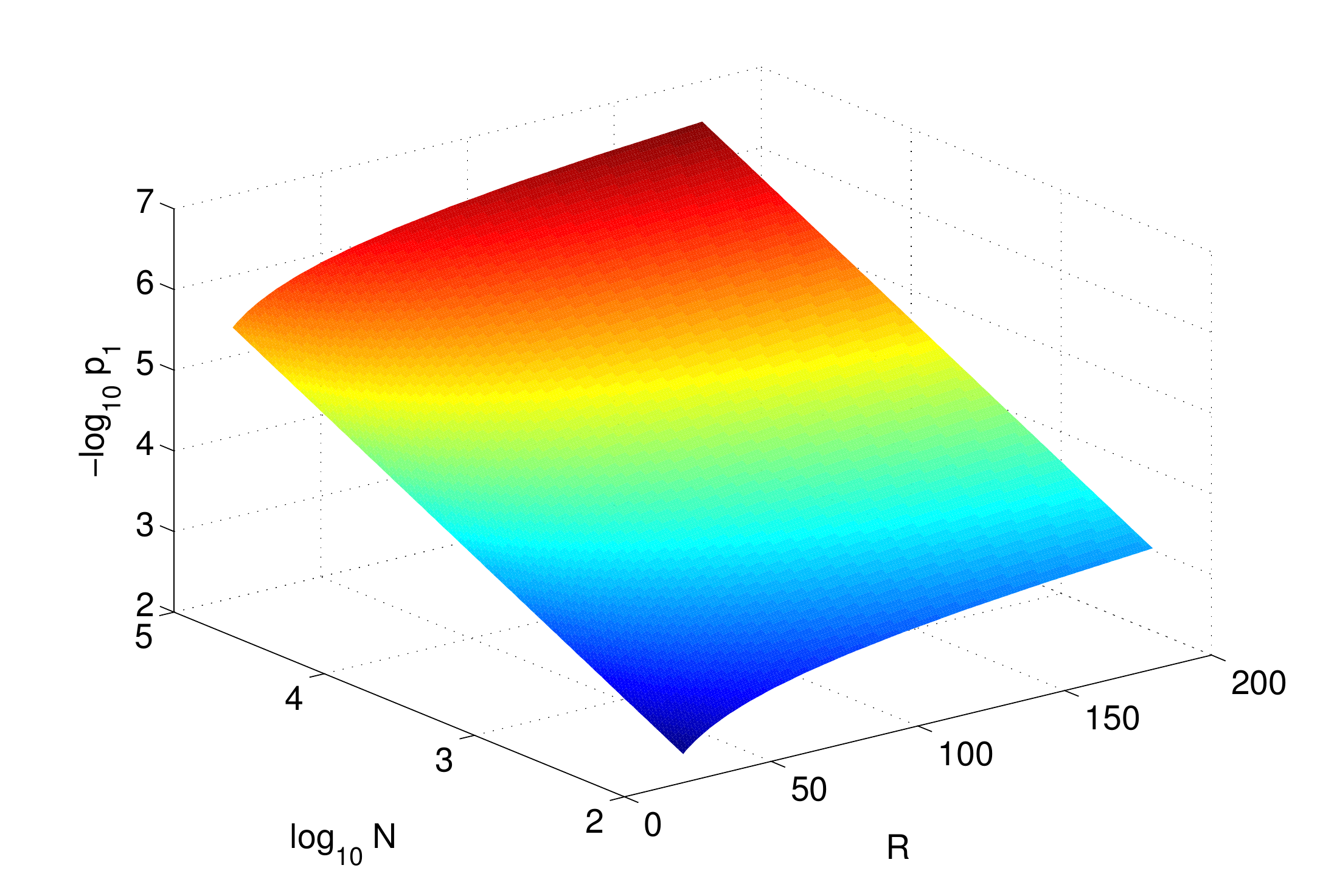}
\caption{Thresholds for switching from simple caching to MBR, i.e. MBR should be used if $p > p_1$. The threshold decreases with both the cost ratio $R$ and the expected number of nodes $N$. Higher z-coordinate on the surface means lower $p_1$.}
\label{scMBRthfig}
\end{figure}

\begin{figure}[tbhp]
\centering \includegraphics[scale=.38]{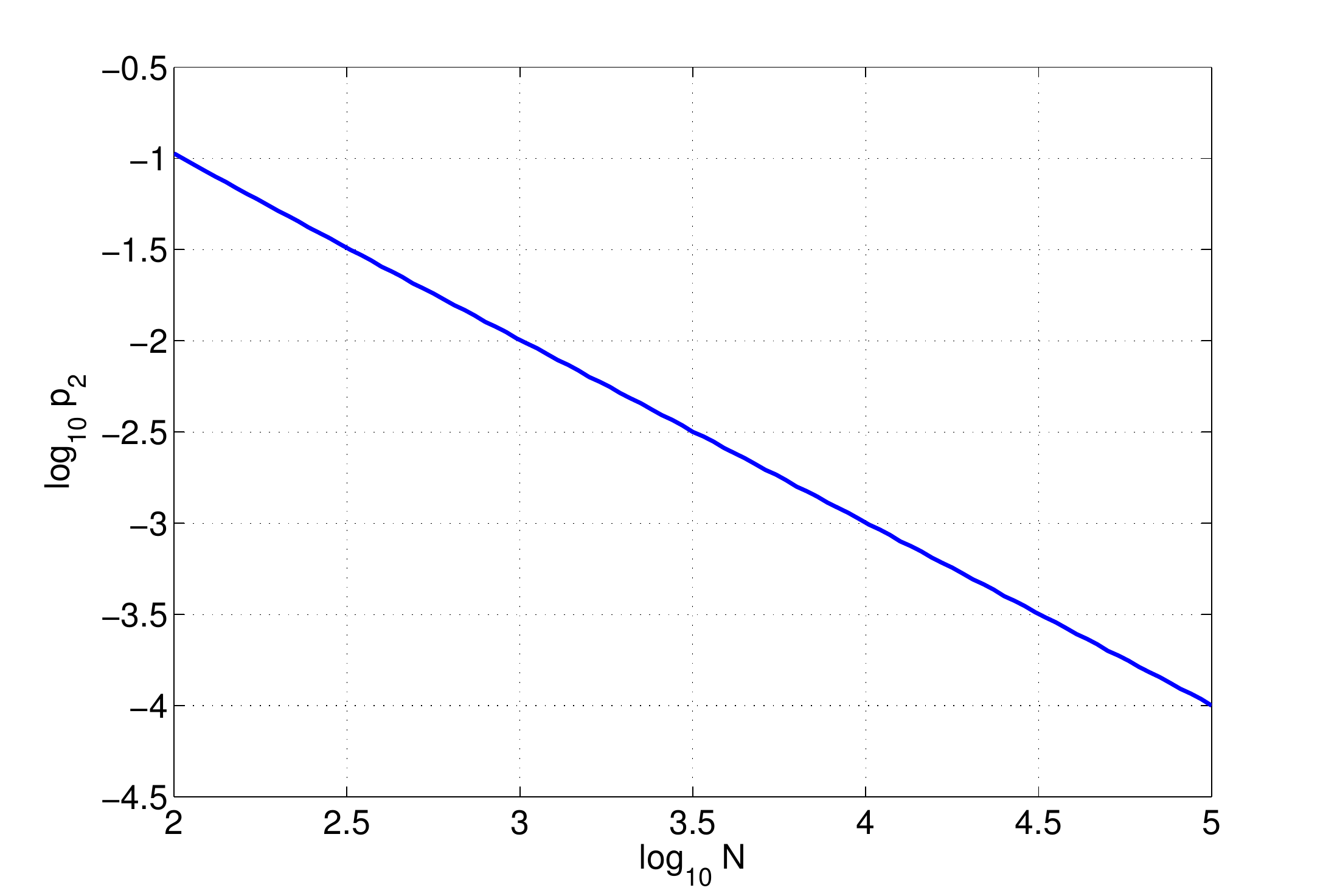}
\caption{Thresholds for switching from MBR to MSR, i.e. MSR should be used if $p > p_2$. We see that $\log_{10}p_2\approx -\log_{10} N+1 \implies p_2\approx \frac{10}{N}$.}
\label{MBRMSRthfig}
\end{figure}

Fig. \ref{scMBRthfig} shows switching thresholds $p_1$ for choosing MBR over simple caching, while Fig. \ref{MBRMSRthfig} shows switching thresholds $p_2$ for choosing MSR over MBR. The switching thresholds are presented for $(R,N)$ parameter pairs with $R\in [20,180]$ and $N\in [10^2,10^5]$. Finding a good curve fit for the surface of Fig. \ref{scMBRthfig} turns out to be rather complicated, and we leave this outside of the scope of this paper. Nonetheless, we find a very simple curve fit for $p_2$ and present it later in this section.

We see that the switching threshold for choosing MBR over simple caching ($p_1$) seems to decrease with both $R$ and $N$. When we increase $R$, contacting the base station becomes more and more expensive. When we increase $N$, the total request rate of the file increases. We see that for high values of $R$ or $N$, it is important to keep the file available on the nodes as we want to avoid having to contact the base station.

As Fig. \ref{MBRMSRthfig} suggests, we can find a very simple approximation for the threshold for choosing MSR over MBR: $p_2\approx \frac{10}{N}$. Here $p_2$ is practically independent of $R$ and only decreases with $N$. This is because it is very unlikely that we need to contact the base station when using either MBR or MSR since $k\ll N$, i.e. it is very unlikely that the number of nodes drops below $k$, which would mean that reconstructing the file is not possible and that we would need to contact the base station. When $N$ increases, so does the expected total request rate of the file, which means that efficient reconstruction becomes increasingly important, and MSR is thus desired.

The threshold for choosing replication over MSR ($p_3$) seems to remain constant at approximately $p_3=0.90$. Again, it is very unlikely that we need to contact the base station when we use either MSR or replication, and the value of $R$ does not matter. As switching from MSR to replication only matters to the nodes that are storing data, and as we keep the number of these nodes constant at $n$, changing $N$ does not affect the decision threshold $p_3$.

It should be noted that all these results hold verbatim only for $n=30$, $k={2,3,...,10}$ and $d=10$. Nevertheless, according to our numerical results, the decision thresholds behave in a similar manner for many other values of $n,k,d$ as well. Therefore, we claim that the behaviour exhibited in these figures also applies to more general settings. However, if the number of storage nodes $n$ is set too high, regenerating codes should not be used. This is because a high value of $n$ incurs a high number of failures and repairs, i.e. a high repair cost. Thus, it is crucial that the system designer chooses well-adjusted values for $n$, $k$, and $d$, which means finding a balance between the number of simultaneous failures that the system needs to withstand, and the expected number of repairs.

\section{Conclusions}\label{conclusionssec}
We have investigated the performance of regenerating codes with many redundant data blocks, caching without redundancy, and replication in a D2D caching system. We have shown that coded storage can offer significant cost savings compared to uncoded storage. We have characterized the decision rules on choosing the optimal method. Coding should only be used if the popularity of the file is between certain thresholds. For very low popularity, no redundancy is required. For very high popularity, replication should be used.

\end{document}